\documentclass[useAMS,usenatbib]{mn2e}

\usepackage{graphicx}
\usepackage{epsfig}  
\usepackage{times}    
\usepackage{subfigure}
\title[An MCMC Fitting Method for Triaxial Dark Matter Haloes]{An MCMC Fitting Method for Triaxial Dark Matter Haloes}
\author[V. L. Corless \& L. J. King]{Virginia L. Corless$^{1}$\thanks{E-mail:
vc258@ast.cam.ac.uk} and Lindsay J. King$^{1}$\\
$^{1}$Institute of Astronomy, University of Cambridge, Madingley Road, Cambridge, United Kingdom}

\begin{document}

\date{2008 May 11}

\pagerange{\pageref{firstpage}--\pageref{lastpage}} \pubyear{2008}

\maketitle

\label{firstpage}

\begin{abstract}
Measuring the 3D distribution of mass on galaxy cluster scales is a crucial test of the $\Lambda$CDM model, providing constraints on the behaviour of dark matter.  Recent work investigating mass distributions of individual galaxy clusters (e.g. Abell 1689) using weak and strong gravitational lensing has revealed potential inconsistencies between the predictions of structure formation models relating halo mass to concentration and those relationships as measured in massive clusters.  However, such analyses employ simple spherical halo models while a growing body of work indicates that triaxial 3D halo structure is both common and important in parameter estimates.  Though lensing is sensitive only to 2D projected structure and is thus incapable of independently constraining 3D models, the very strong assumptions about the symmetry of the lensing halo implied with circular or perturbative elliptical NFW models are not physically motivated and lead to incorrect parameter estimates with significantly underestimated error bars.  We here introduce a Markov Chain Monte Carlo (MCMC) method to fit fully triaxial models to weak lensing data that gives parameter and error estimates that fully incorporate the true uncertainty present in nature.  Using weak lensing data alone, the fits are sensitive to the Bayesian priors on axis ratio; we explore the impact of various general and physically motivated priors, and emphasise the need for future work combining lensing data with other data types to fully constrain the 3D structure of galaxy clusters.  Applying the MCMC triaxial fitting method to a population of NFW triaxial lenses drawn from the shape distribution of structure formation simulations, we find that including triaxiality cannot explain a population of massive, highly concentrated clusters within the framework of $\Lambda$CDM, but easily explains rare cases of apparently massive, highly concentrated, very efficient lensing clusters.  Our MCMC triaxial NFW fitting method is easily expandable to include constraints from additional data types, and its application returns model parameters and errors that more accurately capture the true (and limited) extent of our knowledge of the structure of galaxy cluster lenses.
\end{abstract}

\begin{keywords}
gravitational lensing - cosmology: theory - dark matter - galaxies:clusters: general.
\end{keywords}
\section{Introduction}
Galaxy clusters are ideal laboratories in which to study dark matter, being the most massive bound structures in the universe and dominated by their dark matter component ($\sim90\%$). Constraining the clustering properties of dark matter is crucial for refining structure formation models that predict both the shapes of dark matter halos and their mass function (e.g. \cite{navarro}; \cite{bahcall}; \cite{dahle}).  Several methods are used to measure galaxy cluster dark matter profile shapes and halo masses on a range of scales, including X-ray and Sunyaev-Zeldovich (SZ) studies, dynamical analyses, and gravitational lensing.  However, all of these methods require simplifying assumptions to be made regarding the shape and/or dynamical state of the cluster in order to derive meaningful constraints from available data.  Crucially, while most parametric methods typically assume spherical symmetry of the halo, observed galaxy clusters often exhibit significant projected ellipticity and halos in CDM structure formation simulations (e.g. \cite{bett} (using the Millennium simulation); \cite{shaw}) show significant triaxiality in cluster-scale halos, with axis ratios between minor and major axes as small as 0.4.  Understanding and accurately incorporating the impact of this physical reality on cluster mass and parameter estimates is crucial for accurate comparisons between measured cluster properties and model predictions.

In addition to the determination of masses, most cluster profile fits are carried out in the hope of either supporting or refuting the universality of the NFW profile and thus testing the CDM paradigm.  The NFW profile is typically parameterised by an approximate virial mass $M_{200}$ and a concentration parameter, $C$, and simulations predict a strong correlation between the two.  For a cluster of $M=10^{15}$ M$_{\odot}$, $C \sim 4$.  However, several authors (e.g. \cite{limousin}; \cite{kneib}; \cite{gavazzia}) have recently reported gravitational lensing results in the very low probability tail of the predicted distribution; notably, in a combined weak and strong lensing analysis of Abell 1689, \cite{broadhurst} report a concentration parameter of $C\sim10$, when $C\sim4$ is expected.  While a few such results are not damning, especially given the very complex, likely not relaxed, structures of the lensing halos (see e.g. the work of \cite{lokas} on A1689), it is nonetheless of interest to investigate how possible future discrepancies between observations and the predictions of $\Lambda$CDM should be interpreted.  

Efforts to understand the impacts of triaxiality in gravitational lensing and its potential role in explaining apparent discrepancies with CDM began when \cite{ogurib} applied a fully triaxial NFW model to the shear map of Abell 1689 to find that it is consistent with $6\%$ of cluster-scale halos. These continued as \cite{gavazzib} showed that a triaxial NFW can reconcile parameter values derived from observations of the cluster MS2137-23 to predictions from N-body simulations.

More generally, \cite{corl} demonstrated in the weak lensing regime that neglecting halo triaxiality in parameterised fits of NFW models to NFW haloes with axis ratios significantly less than one can lead to over- and underestimates of up to 50$\%$ and a factor of 2 in halo mass and concentration, respectively.  While extreme cases of triaxiality are rare, such haloes can be much more efficient lenses than their more spherical counterparts, especially when in configurations that hide most of the triaxial shape along the line of sight.  Further, even haloes with less extreme axis ratios are inaccurately fit by spherical models.  We expect the vast majority of galaxy cluster scale dark matter haloes, and so it is important to include this expectation in model fitting.

In the past, triaxial models have not generally been fit to lensing data because they cannot be well-constrained.  A simple NFW triaxial halo model has six free parameters (concentration, mass, two axis ratios, and two orientation angles) as opposed to the two parameters of the spherical model (mass and concentration only), and even combined weak and strong lensing data is not enough to meaningfully constrain so many parameters.  This is sometimes due to the limited depth of currently available observational data, but more importantly to the intrinsic limitations of lensing.  Because lensing is affected only by the the projected surface density and shear of the underlying mass distribution, it is inherently impossible to fully constrain a three-dimensional structure without imposing strong priors on the shape of the halo or supplementing lensing data with other data types more sensitive to line-of-sight halo structures, such as dynamical studies.

Despite these difficulties, since triaxiality has been convincingly demonstrated to be an important factor in model parameter estimation in lensing analyses, it is necessary to find a way to directly include it in NFW fits to galaxy clusters. For individual clusters it is crucial, if claims regarding the validity of the $\Lambda$CDM paradigm based on NFW parameter estimates are to be meaningfully evaluated.  It is also of importance across populations to obtain more accurate distributions of galaxy cluster parameters, and especially in measuring the galaxy cluster mass function -- if the mass function is signficantly sloped as expected, even the best-case scenario of a symmetric scatter of mass estimates due to neglected triaxiality would lead to an asymmetric shift in the calculated mass function, as there are more low mass halos to shift up in mass than there are 
high mass halos to shift down.

To that end, we present here a Bayesian Markov Chain Monte Carlo method to fit fully triaxial NFW halos to weak lensing data. This method flexibly 
combines weak lensing data with prior probability functions on the model halo parameters to return parameter and error estimates that reflect the 
true uncertainties of the problem. Importantly, it also allows for the straightforward and statistically-robust inclusion of additional 
constraints from other data sources such as observations from SZ, X-ray and spectroscopic surveys.

\section{Lensing by Triaxial NFW Halos}
\subsection{Weak Lensing Background}\label{subsec:wl}
Weak lensing distorts the shapes and number densities of background galaxies.  The shape and orientation of a background galaxy can be described by a complex ellipticity $\epsilon^s$, with modulus $|\epsilon^s|=(1-b/a)/(1+b/a)$, where $b/a$ is the minor:major axis ratio, and a phase that is twice the position angle $\phi$, $\epsilon^s=|\epsilon^s|e^{2i\phi}$.  The galaxy's shape is distorted by the weak lensing reduced shear, $g=\gamma/(1-\kappa)$, where $\gamma$ is the lensing shear and $\kappa$ the convergence, such that the ellipticity of the lensed galaxy $\epsilon$ becomes
\begin{equation}\epsilon = \frac{\epsilon^s + g}{1 + g^{\ast}\epsilon^s} \approx \epsilon^s + \gamma\label{eq:lens}\end{equation}
in the limit of weak deflections.  The distributions of ellipticities for the lensed and unlensed populations are related by 
\begin{equation}p_{\epsilon} = p_{\epsilon^s}\left|\frac{d^2\epsilon^s}{d^2\epsilon}\right|;\end{equation}
assuming a zero-mean unlensed population, the expectation values for the lensed ellipticity on a piece of sky is $<\epsilon> = g \approx \gamma$.  This is the basis for weak lensing analysis in which the shapes of images are measured to estimate the shear profile generated by an astronomical lens.


Lensing also changes the number counts of galaxies on the sky via competing effects; some faint sources in highly magnified regions are made brighter and pushed above the flux limit of the observation, but those same regions are stretched by the lensing across a larger patch of sky and so the number density of sources is reduced.  Thus the number of sources in the lensed sky $n$ is related to that in the unlensed background $n_0$ and the slope of the number counts of sources at a given flux limit $\alpha$ by $n=n_0\mu^{\alpha-1}$, where $\mu$ is the lensing magnification $\mu^{-1}=(1-\kappa)^2 - |\gamma|^2$.  A full description of these effects is given in \cite{canizares}.

We now describe the characteristic behaviour of the convergence $\kappa$ and shear $\gamma$ of triaxial NFW halos.

\subsection{Triaxial NFW}
A full parameterisation for a triaxial NFW halo is given by \cite{jing} (herein JS02).  They generalise the spherical NFW profile to obtain a 
density profile
\begin{equation}
\rho(R) = \frac{\delta_c \rho_c(z)}{R/R_s(1 + R/R_s)^2}
\label{eq:3axrho}
\end{equation}
where $\delta_c$ is the characteristic overdensity of the halo, $\rho_c$ the critical density of the Universe at the redshift $z$ of the cluster, $R_s$ a scale radius, $R$ a triaxial radius 
\begin{equation}
R^2 = \frac{X^2}{a^2} + \frac{Y^2}{b^2} + \frac{Z^2}{c^2},\textrm{         }(a\leq  b \leq  c = 1),\label{eq:3axR}\end{equation}
and $a/c$ and $b/c$ the minor:major and intermediate:major axis ratios, respectively.  In a  different choice from JS02 we define a triaxial virial radius $R_{200}$ such that the mean density contained within an ellipsoid of semi-major axis $R_{200}$ is $200\rho_c$ such that the concentration is
\begin{equation}C = \frac{R_{200}}{R_s},\label{eq:3axC}\end{equation}
the characteristic overdensity is
\begin{equation}\delta_c = \frac{200}{3} \frac{C^3}{\log (1+C) - \frac{C}{1 + C}},\label{eq:3axdelta}\end{equation}
the same as for a spherical NFW profile, and the virial mass is
\begin{equation}M_{200} = \frac{800\pi}{3}abR_{200}^3\rho_c.\label{eq:3axM200}\end{equation}

In past lensing studies, an effective spherical virial mass and concentration have often been employed instead of these triaxial definitions, in part to allow more straightforward comparison to populations of structures extracted from simulations using spherical halo boundaries.  The effective spherical virial radius is defined as the radius $r_{200}$ at which the mean density within a sphere of that radius is 200 times the critical density, and the effective spherical virial mass the mass within that sphere: 
\begin{equation} m_{200} = (800\pi /3) r_{200}^3\rho_c.\end{equation}
We further define the effective spherical concentration $C_{sph}$ of a triaxial halo as the ratio of the effective spherical virial radius to the geometric mean of the triaxial scale radii $r_s = R_s(abc)^{1/3}$:
\begin{equation}C_{sph} = r_{200}/r_s.\end{equation}
We test the performance of our fitting method in recovering these effective values for the purposes of comparison with previous work; however, we stress that a fully triaxial definition of virial mass better represents the true shape of dark matter halos, and is well motivated by ellipsoidal collapse models that predict asymmetric collapse to stop at the same enclosed density as does spherical collapse (\cite{s2}).  We therefore prefer the fully triaxial parameterisation in this and future work as the more physically motivated choice.  In practice, the effective spherical mass and concentration are always less than and vary only moderately from their fully triaxial counterparts, by up to $10\%$ for very triaxial halos and most often far less.

The triaxial halo is oriented with respect to the observer's line of sight by angles $\theta$ and $\phi$; randomly oriented halos are distributed uniformly in $\phi$ and sin$\theta$.  A more detailed description of this parameterisation and its benefits is given in \cite{corl}.

\begin{figure}
\epsfig{file=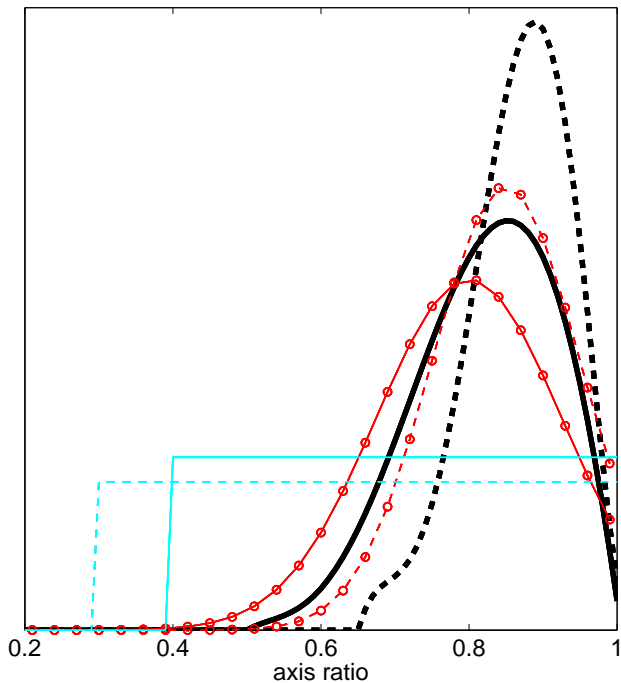,scale=0.5}
 \caption{The prior probability functions for axis ratios $b$ (solid lines) and $a/b$ (dashed lines).  The {\bf bold} lines show the Shaw prior, the o-dotted lines the Axis Gaussian prior generalised from Bett et al., and the lightest lines the Flat prior.}
 \label{fig:plot1}
\end{figure}

The full derivation of the lensing properties of a triaxial halo is given by \cite{oguria} (herein OLS), and we summarise and extend some of that 
work in Appendix \ref{sec:appa}.

\begin{figure}
\includegraphics[width=80mm, height=80mm]{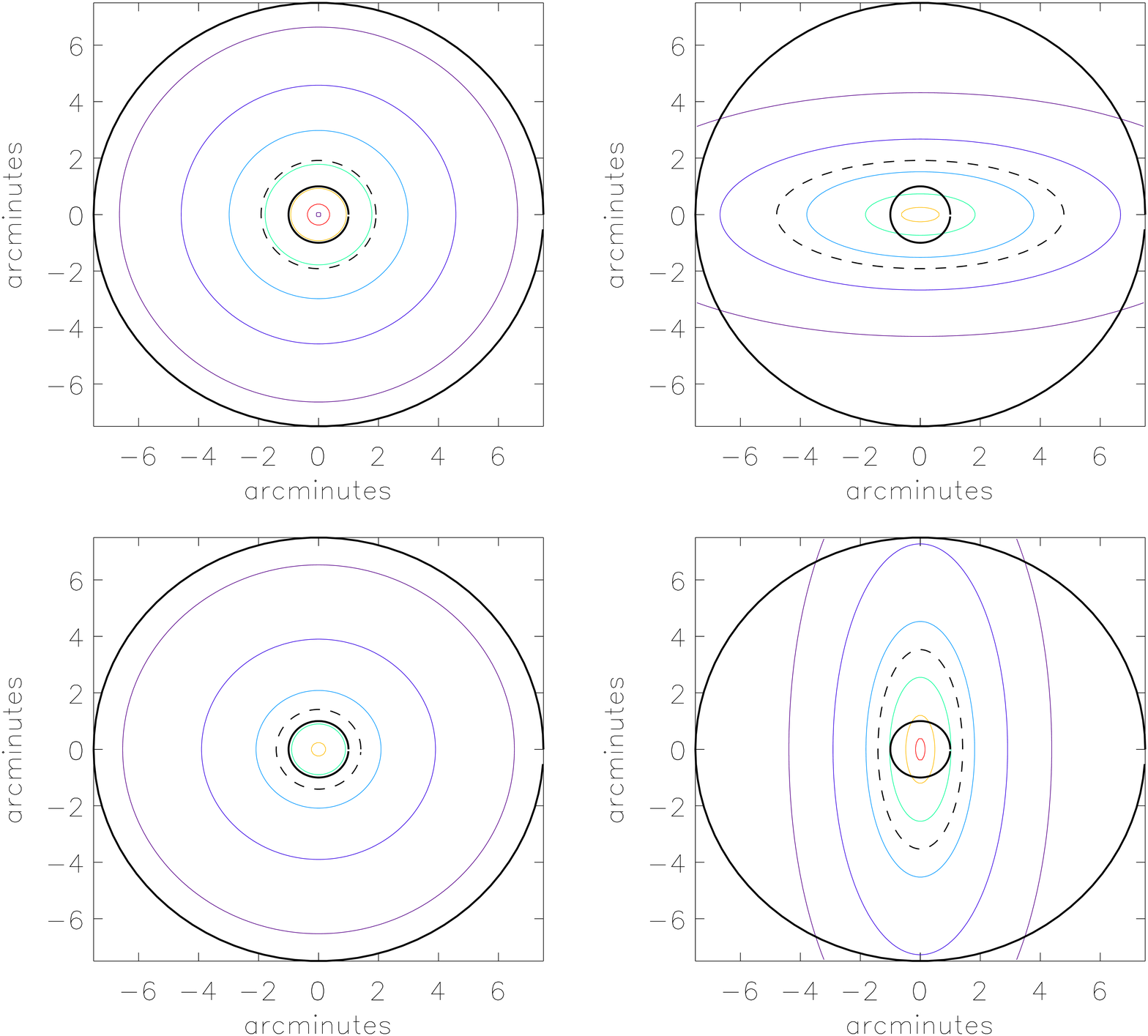}
 \caption{The top \{lower\} panels show  isoconvergence contours for a prolate \{oblate\} halo of $M_{200}=10^{15}$ M$_{\odot}$, $C=4$, with axis ratios $a=b=0.4$ \{$a=0.4, b=1.0$\}; the left-hand panel shows the halo oriented with the odd axis along the line of sight (LoS), the right-hand panel shows the halo with odd axis in the plane of the sky (Plane).  The thick solid lines show the limits of the aperture from which weak lensing data is taken, and the dashed line shows the $R_s$ ellipse for each projection (note that $R_s$ is scaled by the axis ratio in each direction, so that when looking at the minor axis of a triaxial halo, the apparent scale radius is $a$ times the $R_s$ value of the halo).  The lowest contour corresponds to $\kappa = 0.02$ and subsequent contours each increase in $\kappa$ by a factor of 2.}
 \label{fig:plot2}
\end{figure}

\begin{figure*}
\epsfig{file=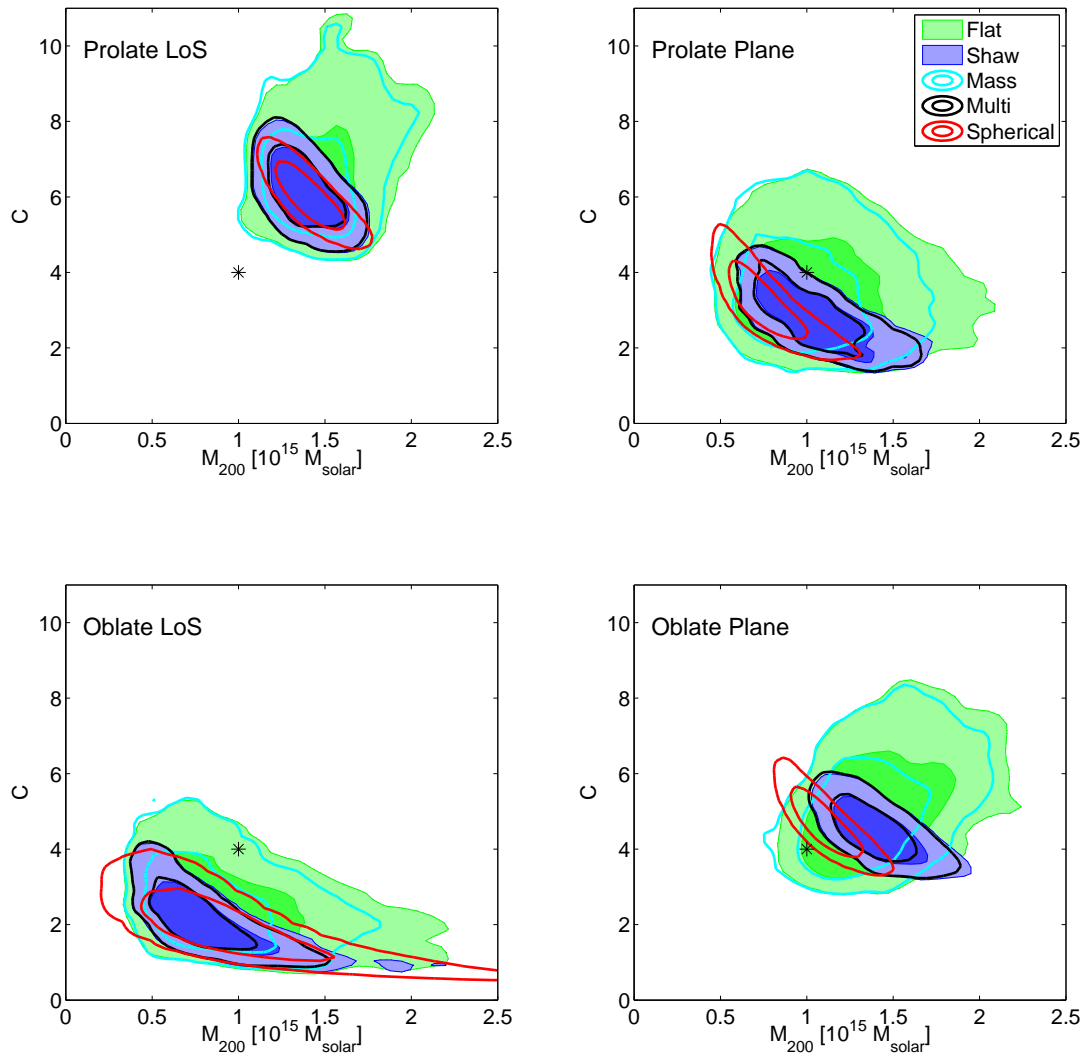,scale=0.5}
\caption{$M_{200}-C$ confidence contours at 68$\%$ and 95$\%$ for the triaxial NFW fit to four highly triaxial lenses under various priors.  These are the four extreme cases noted in our first triaxiality paper: symmetric oblate and prolate halos with $M_{200}=10^{15}$ M$_{\odot}$ and $C=4$ and axis ratios $\{a=0.4, b=1.0\}$ and $\{a=b=0.4\}$, respectively, aligned in Line-of-Sight and Plane of the Sky orientations.  The black stars show the true parameters of the underlying lens.}
\label{fig:plot3}
\end{figure*}

\begin{figure*}
\epsfig{file=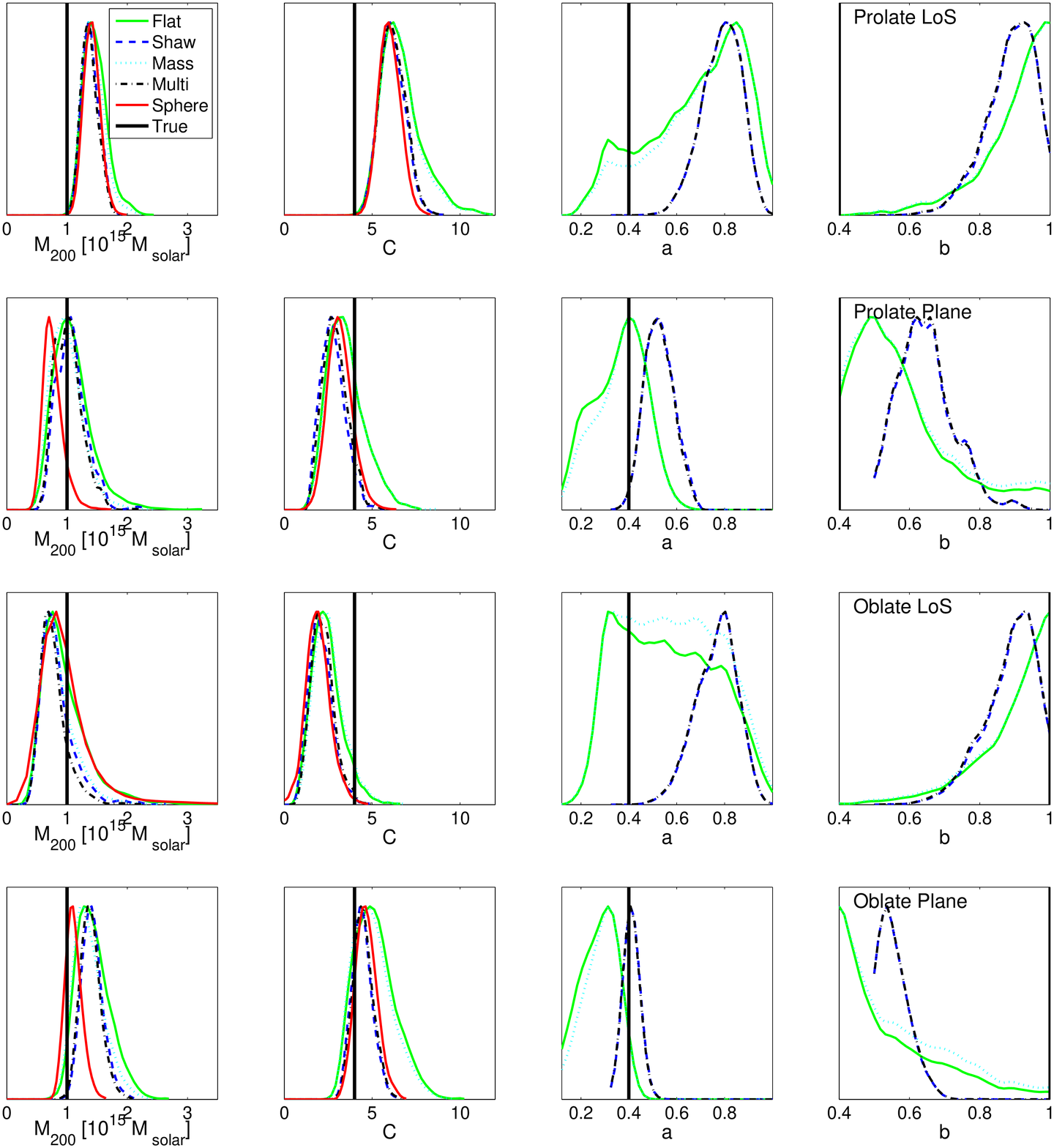,scale=0.4}
\caption{1D parameter distributions for the triaxial NFW fitted to weak lensing by four highly triaxial halos under various priors.  These are for the four extreme cases noted in our first triaxiality paper: symmetric oblate and prolate halos with $M_{200}=10^{15}$ M$_{\odot}$ and $C=4$ and axis ratios $\{a=0.4, b=1.0\}$ and $\{a=b=0.4\}$, respectively, aligned in Line-of-Sight and Plane of the Sky orientations.  The thick black lines show the true parameters of the underlying lens.}
\label{fig:plot4}
\end{figure*}

\begin{figure*}
\epsfig{file=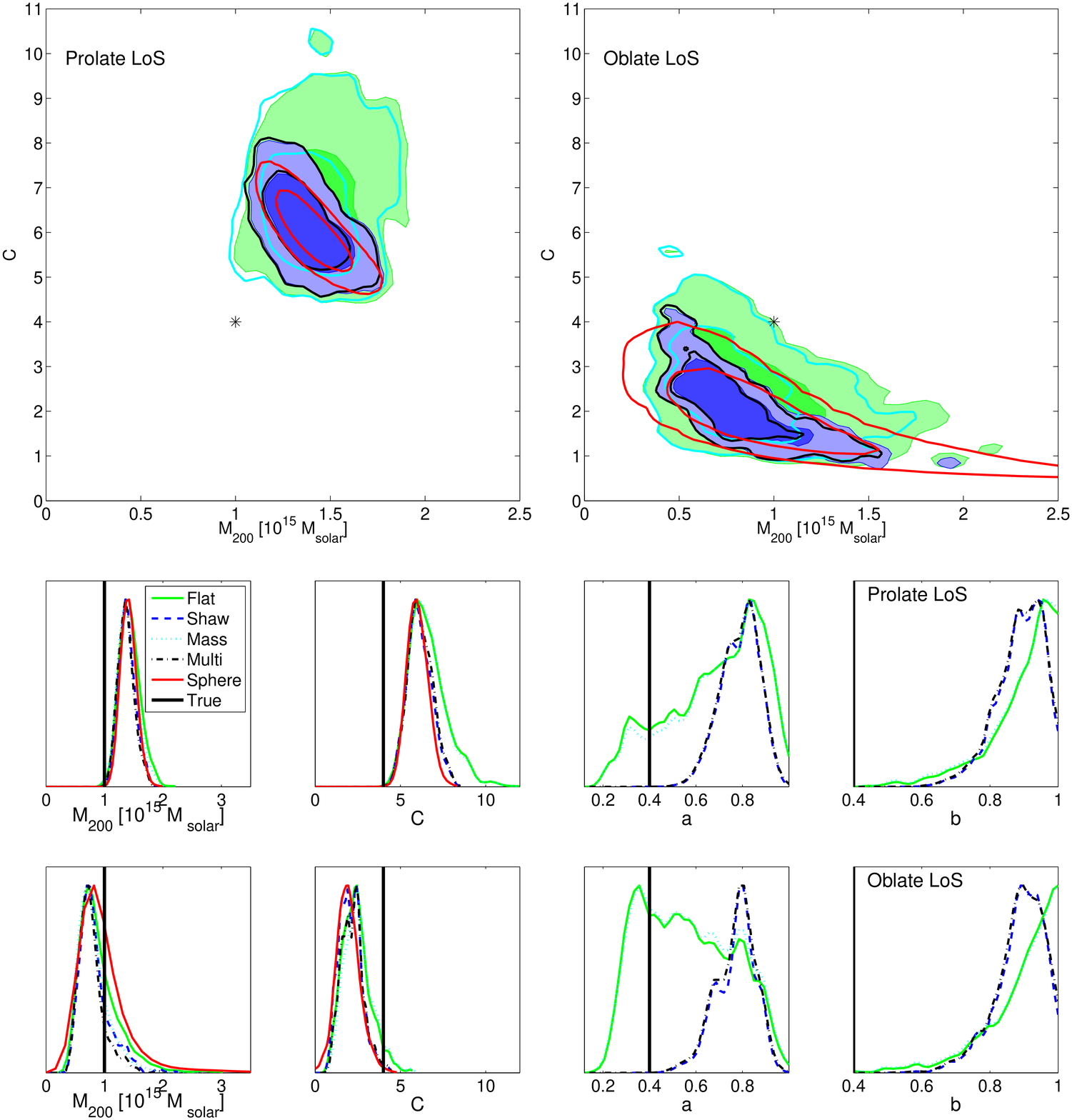,scale=0.4}
\caption{The top panels show  the $m_{200}-C_{sph}$ confidence contours at 68$\%$ and 95$\%$ for the effective spherical parameterisation of the triaxial NFW fit to the Line-of-Sight oriented highly triaxial prolate and oblate lenses of Figure \ref{fig:plot3} under various priors.  The black stars show the true effective spherical parameters of the underlying lens.  The bottom panels plot the 1D effective spherical parameter distributions for the same lenses and priors, for comparison with \ref{fig:plot4}.  The thick black lines show the true effective spherical parameters of the underlying lens.}
\label{fig:plot5}
\end{figure*}



\subsection{Weak Lensing Simulations}\label{sec:simulations}

The main body of simulations is carried out for a field $7.5'$ in radius, with a background source density $n_0=30/$arcminute$^2$ (Poisson noise is accounted for), typical of ground-based observations.  Although wide-field imaging of clusters with fields of half a degree is routinely possible, there are arguments that errors due to large-scale structure along the line-of-sight become more important as the shear due to the cluster itself diminishes with distance from the centre (e.g. \cite{hoekstra}).  In any case, the general trends will hold for larger fields. A catalogue of randomly positioned and oriented galaxies with intrinsic shapes $\epsilon^s$ drawn from a Gaussian distribution with dispersion $\sigma_{\epsilon}=\sqrt{\sigma_{\epsilon 1}^2 + \sigma_{\epsilon 2}^2} = 0.2$ in the modulus $|\epsilon^s|$ is placed at redshift $z=1$.  This catalogue of background galaxies is lensed through a model lens of choice placed at redshift $z=0.18$ (the redshift of Abell 1689), at which the width of the field is $\sim 1900$ kpc/h.  Thus our choice to place all sources on a sheet at $z=1$ is justified by the low redshift of our fiducial lens; only for higher redshift lenses that are in the heart of the redshift distribution is the distribution of source redshifts important \citep{seitz}.  The background galaxies are lensed according to Equation \ref{eq:lens} and the number counts are reduced as prescribed in section~\ref{subsec:wl}, taking the slope of the source number counts in flux to be ${\rm d log}N/{\rm d log}S=\alpha=0.5$ (corresponding to a slope of 0.2 in magnitude as in \cite{fort}).  Galaxies located within $1'$ of the cluster centre are removed from the analysis to avoid the strong lensing regime at the centre of the cluster (in any case background galaxies near the cluster centre would be mostly obscured by cluster members in observations).  Throughout we assume a concordance cosmology with $\Omega_m=0.3$, $H_0 = 70$ km s$^{-1}$ Mpc$^{-1}$, and a cosmological constant $\Omega_{\Lambda} = 0.7$, and a typical massive cluster of triaxial $M_{200}=10^{15}$ M$_{\odot}$ and $C=4$.  At the redshift of our lens $z=0.18$, 1 arcminute corresponds to $\sim 127$ kpc/$h$.

\begin{figure*}
\epsfig{file=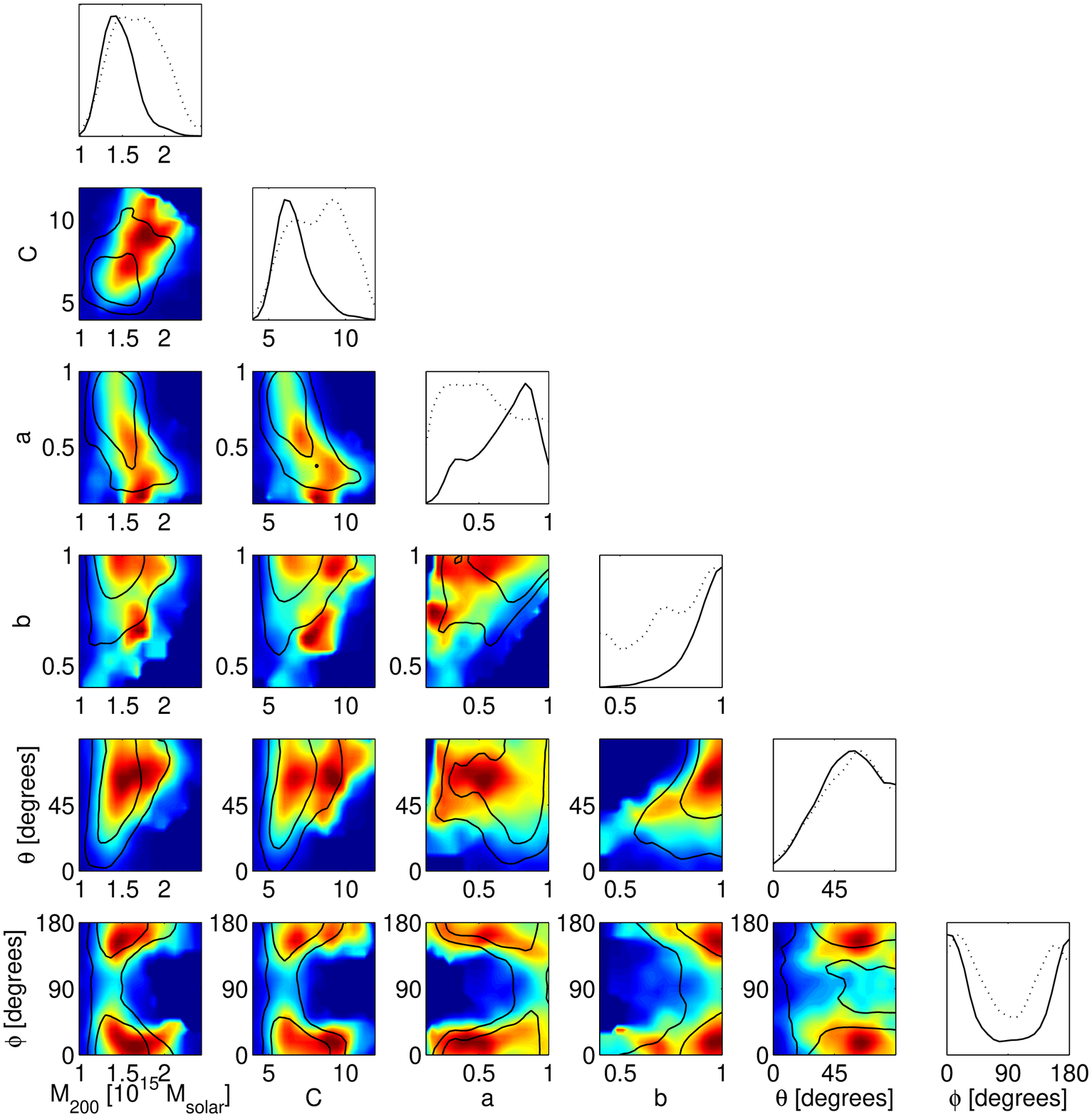,scale=0.6}
\caption{The full posterior probability distribution for a triaxial NFW fitted to weak lensing by an extreme prolate halo {a=b=0.4} in a Line of Sight orientation, under a Flat prior.  The black contours give the 68$\%$ and 95$\%$ confidence contours of the marginalized posterior probability distribution, and the shading shows the marginalized likelihood values, smoothed over 20 bins (rather than 40 as in other figures) for increased clarity.  The solid (dotted) lines in the unshaded distributions at the end of each row give the 1-parameter marginalized probability (likelihood) distributions.}
\label{fig:plot6}
\end{figure*}

\begin{figure*}
\epsfig{file=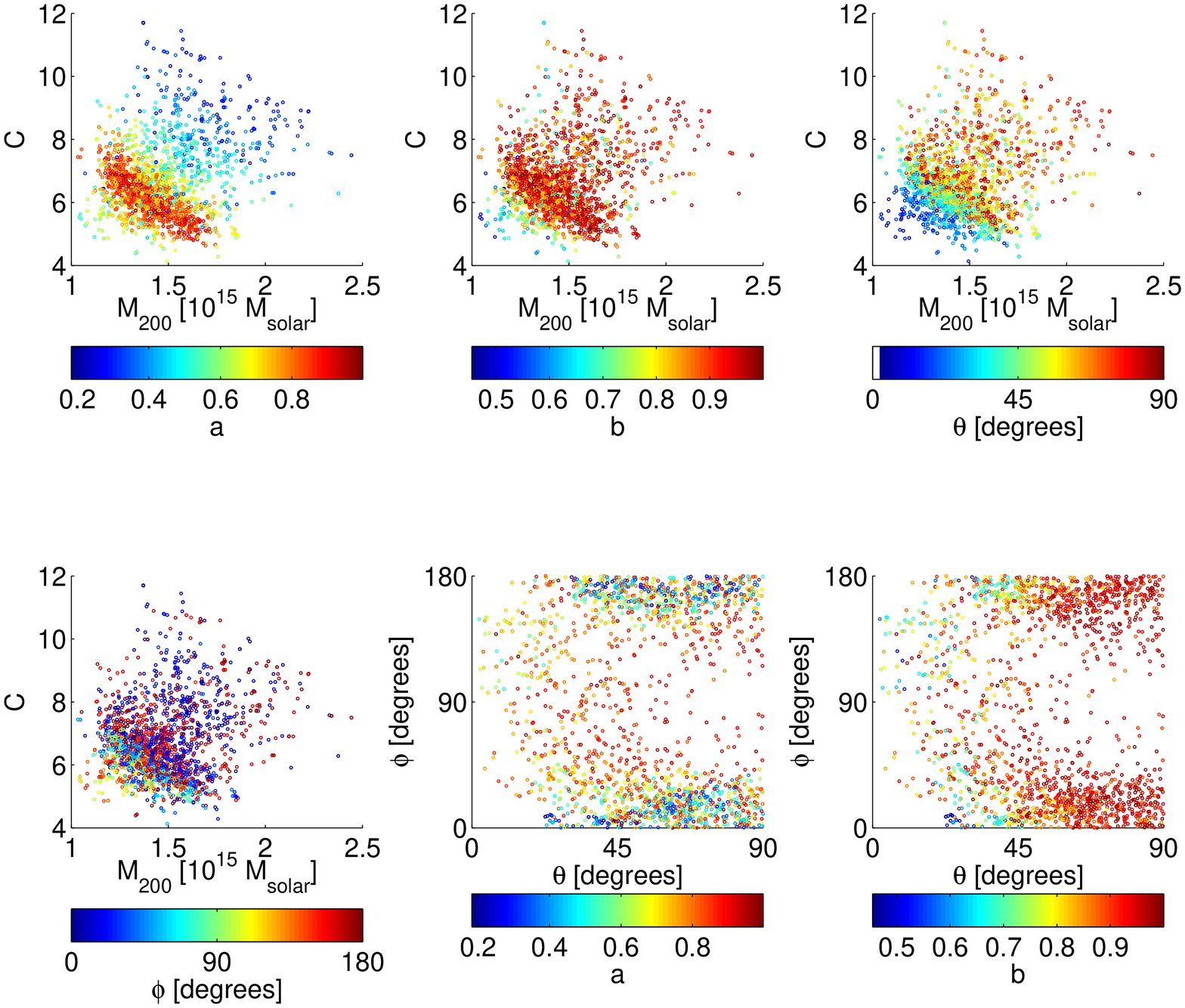,scale=0.6}
\caption{$M_{200}-C$ contours and $\theta-\phi$ contours shaded by the axis ratios and, for the first pairing, orientation angles for a triaxial NFW fitted to weak lensing by an extreme prolate halo {a=b=0.4} in a Line of Sight orientation, under a Flat prior.}
\label{fig:plot7}
\end{figure*}

\section{Markov Chain Monte Carlo Fitting}
Markov Chain Monte Carlo (MCMC) methods are hugely valuable in under-constrained or highly-degenerate fitting problems; they are employed widely in CMB analyses and are rapidly growing in popularity in lensing work (see e.g. \cite{marshall}, \cite{jullo}, \cite{corlb}).  This family of methods employ a ``guided'' random walk that returns a sample of points representative of the posterior probability distribution; the probability of a certain region of parameter space containing the true model is directly proportional to the density of points sampled in that region.  From the distribution of sample points the full posterior probability distribution is obtained, which is easily and directly marginalized over to obtain fully marginalized mean most-probable parameter estimates for all parameters.  These most-probable parameter values reflect the full shape of the posterior probability distribution without assumption about the shape of the error distribution, and taking full account of the prior.  Such methods have exploded in popularity recently, and there are several excellent references describing the method in detail, e.g. \cite{lewis}, \cite{mackay}.  Simply put, the sampler functions by stepping through parameter space by taking random steps governed by a transfer function, usually a simple $n$-D Gaussian, where $n$ is the number of parameters of the fitted model.  If the randomly-chosen next step is to a point of higher probability than the current position, the step is taken.  If the next step is to a point of lower probability, the step is taken with probability $p({\rm new})/p({\rm current})$.  Thus, the MCMC sampler spends most of its time in high probability regions, but can move ``downhill'' to regions of lower probability in order to explore the entire space and sample all regions of high probability.  Crucially, this method is able to return a true representation of the full posterior probability distribution, regardless of the complexities of that probability distribution, e.g., tight degeneracies, multiple islands of high probability, or a very flat distribution due to poor constraints from the data.  

Fitting triaxial models with lensing data is an intrinsically under-constrained problem, as lensing on cluster scales can {\it never} give complete information about the 3-D triaxial structure because all lensing behaviour is determined only by the 2-D projected mass density and potential.  
With such weak constraints, a maximum likelihood approach is both impractical and of very limited scientific value; the posterior distribution will be very flat with significant degeneracies giving poorly constrained maximum likelihood values.  By exploring the full posterior probability distribution, this MCMC method allows the derivation of parameter estimates (and their accompanying errors) that account for the true uncertainties when fitting parametric models to lensing data.

To implement the method, we must define the posterior probability function, defined in Bayesian statistics as
\begin{equation}p(\pi|\theta) = \frac{p(\theta|\pi)p(\pi)}{p(\theta)}\end{equation}
where $p(\theta|\pi)$ is the likelihood $\mathcal{L}$ of the data given the model parameters (the standard likelihood), $p(\pi)$ is the prior probability distribution for the model parameters (e.g. a distribution of axis ratios drawn from simulations), and $p(\theta)$ is a normalising factor called the {\it evidence}, of great value in comparing models of different classes and parameter types, but expensive to calculate and unnecessary for the accurate exploration of the posterior distribution.  
The assignment of priors is often a controversial exercise; we will return to this question since priors acquire an increased significance in under-constrained problems such as ours.   We define the log-likelihood function in the standard manner for weak lensing following \cite{schneider} and \cite{king}
\begin{equation}\ell_{\gamma} = -\ln \mathcal{L} = -\sum_{i=1}^{n_{\gamma}}\ln  p_{\epsilon}(\epsilon_i|g(\vec{ \mathcal{\theta}_i}; \Pi)).\end{equation}
where the reduced shear $g$ is calculated using the triaxial convergence and shear of Equations \ref{eq:gammakappa}, and $\Pi$ is a six-element vector of the parameters defining the model: triaxial virial mass $M_{200}$, concentration $C$, minor axis ratio $a$, intermediate axis ratio $b$, and two orientation angles $\theta$ and $\phi$.  

In our MCMC sampler we employ a 6D two-sided Gaussian transfer function, and use the covariance matrix of an early run to sample in an optimised basis aligned with the degeneracies of the posterior.  We tune the step sizes of the sampler to achieve an average acceptance rate of 1/3 in each basis direction, run three independent MCMC chains, started at randomly chosen positions in parameter space, for each lens system, and sample the distribution space until the standard var(chain mean)/mean(chain var) indicator is less than 0.2, indicating chain convergence.  We utilise the GetDist package from the standard CosmoMC (\citealt{lewis}) distribution to calculate convergence statistics, parameter contours, and marginalized parameter estimates.  We employ 40 bins in the Gaussian smoothing of the contours, chosen because, for the spherical case, this smoothing length best matches the contours obtained using this MCMC method to those of a standard $\chi^2$ approach.  We note that while the angles are defined over a range $0<\theta <\pi$ and $0< \phi < 2\pi$, because of the elliptical nature of the projected density contours, they give rise to unique lensing profiles only over the range of $0<\theta <\pi/2$ and $0< \phi < \pi$.

\subsection{Choice of Priors}
As noted above, the choice of priors used on the parameters of the triaxial NFW model is very important because of the inherently under-constrained nature of the problem.  The standard analysis method using a spherical NFW is equivalent to putting $\delta$-function priors on both axis ratios at $a=b=1$.  This is clearly a very strong prior, and its application to lensing halos with significant triaxial structure was shown in CK07 to lead to errors in mass and concentration estimates of factors of up to 2, in some cases in the same direction (i.e. increased mass {\it and} concentration), causing potentially misleading disagreements with the inverse $C-M$ relationship predicted in $\Lambda$CDM simulations.  

The opposite of this strong prior is a very weak flat prior on both axis ratios truncated at some small value.  Here, as a representative of this class of prior, we impose 
\begin{eqnarray}
0.4\leq &b&\leq 1.0\textrm{   and}\nonumber\\
0.3\leq &a/b&\leq 1.0,\end{eqnarray}
giving an effective prior on the minor axis ratio $0.12\leq  a\leq 1.0$, and refer to this herein as the Flat prior.  This is a very general and weak prior, but, as with the spherical case, does not reflect our true prior beliefs about the axis ratio distribution of triaxial halos.  We do not use the weakest possible prior, $0.0 < a=b \leq  1.0$, because we find that allowing unphysically tiny axis ratios makes convergence of the MCMC runs very difficult.  However, the Flat prior could be extended to even lower axis ratios and the length of the MCMC runs extended, if the problem at hand demands it.

An intermediate choice is to use distributions from CDM simulations, such as those of \cite{shaw}, as the prior distribution on axis ratios.  We fit polynomial functions to the distributions of $b$ and $a/b$, the intermediate:major axis ratio and the minor:intermediate axis ratio -- the exact functions used are give in Appendix \ref{sec:appb} -- and the distributions as used are plotted in Figure~\ref{fig:plot1}.  Note that the distribution for $b$ peaks at about 0.8 and that for $a/b$ peaks at about 0.9, giving a population, as found in several other simulations as well (see e.g. \cite{bett}), skewed slightly toward prolate halos.  We will call this the Shaw prior herein.  

Additionally, one might wish to place a prior on the mass of the lensing halo, based on a simulated or observed mass function.  For simplicity, we adopt a simple exponential prior 
\begin{equation}p(M_{200}) \propto \exp\left(-M_{200}/10^{15} M_{\odot}\right)\end{equation}
to understand the strength of this class of prior compared to others, and refer to this as the Mass prior.

There are many other possible choices of prior, including a more general prior on the axis ratios tending towards less extreme values than the Flat prior, a prior based on the $C-M$ relationship predicted by $\Lambda$CDM, a prior taking into account selection biases due to lensing efficiency (some halo shapes and orientations -- i.e. prolate line of sight -- are much better lenses than others), or priors based on knowledge about a specific lensing system gained from outside data or unusual characteristics of the system. In addition, many of these priors may be used in combination.  Exploring the impact of priors is relatively computationally cheap because they can often be applied to existing MCMC chains that have been run under more general priors; a post-processing conversion can be performed as long as everywhere the new prior is non-zero there was non-negligible probability under the original prior.  Thus, the Shaw and Mass priors can usually be applied after the fact to runs under the Flat prior.  By contrast, the Spherical prior cannot, as there is too little relative probability and therefore far too few samples at $a=b=1$ under the Flat prior.

The choice of prior is always a balance between imposing enough external constraints to gain usefully specific parameter estimates without falsely over-constraining the problem with priors too strong or simply wrong for the problem at hand.  We explore the behaviour of the triaxial NFW fits under several of these priors in the following section, and return to this important question in the Discussion.
\begin{figure}
\epsfig{file=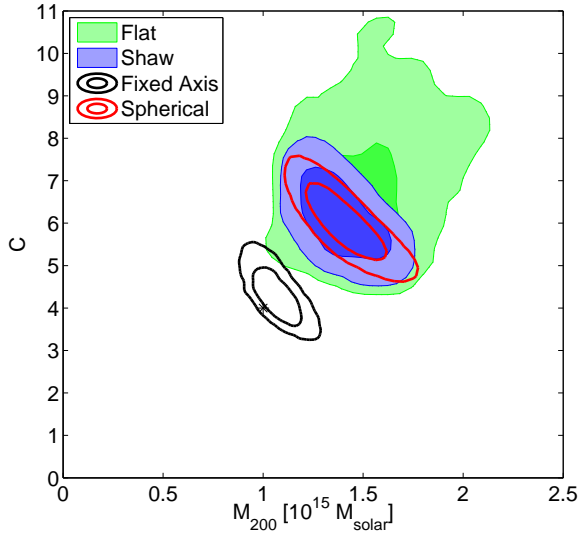,scale=0.5}
\caption{$M_{200}-C$ contours for a triaxial NFW fitted to weak lensing by an extreme prolate halo {a=b=0.4} in a line of sight orientation under several priors.  The black star gives the true parameters of the underlying lens.  The black contours are the result for a triaxial NFW fit with both axis ratios fixed at the the true values {$a=b=0.4$}; as expected, they enclose the true parameter values for mass and concentration.}
\label{fig:plot8}
\end{figure}

\section{Results and Analysis}

\subsection{Fitting to highly triaxial halos}\label{ssec:ex}
To begin to understand both the need for this triaxial fitting method and its behaviour under different priors, we examine four cases of highly triaxial halos studied in CK07.  They are prolate and oblate halos of $M_{200} = 10^{15}M_{\odot}$ and $C=4$ with axis ratios ($a=b=0.4$) and ($a=0.4; b=1.0$) respectively, with effective spherical virial masses and concentrations ($m_{200} = 0.90\times 10^{15}M_{\odot}$, $C_{sph} = 3.86$) and ($m_{200} = 0.91\times 10^{15}M_{\odot}$, $C_{sph} = 3.87$) oriented in Line of Sight and Plane of the Sky orientations; Figure~\ref{fig:plot2} shows the convergence contours for these four lensing configurations.  CK07 showed that these four lenses, if fit with spherical models using a standard maximum likelihood technique give mean parameter values that differ significantly from the true values of the lens:
\begin{itemize}
\item Prolate LoS: $M_{200} = 1.34\times 10^{15} M_{\odot}; C=6.0$
\item Prolate Plane: $M_{200} = 0.75\times 10^{15} M_{\odot}; C=3.4$
\item Oblate LoS: $M_{200} = 0.72\times 10^{15} M_{\odot}; C=2.7$
\item Oblate Plane: $M_{200} = 1.05\times 10^{15} M_{\odot}; C=5.0$.
\end{itemize}

Figure~\ref{fig:plot3} shows the resulting $M_{200} - C$ contours fitting a triaxial NFW to a single lensing realization through each of these four cases applying a Flat, Shaw, Mass, joint Shaw and Mass (Multi), and Spherical prior. Figure \ref{fig:plot4} plots the 1d marginalized parameter distributions. As expected the broader the prior, the broader the error contours.  In these extreme cases the Flat prior performs best; this is unsurprising in comparison to the Shaw prior, which has low probabilities of axis ratios so extreme as those of these halos, and certainly in comparison to the spherical prior.  The Mass prior proves weak, reducing the contours only slightly from the Flat prior case, in the expected direction away from high mass solutions.

The four halos show trends consistent with those expected from previous work in their relative masses and concentrations.  For comparison, Figure~\ref{fig:plot5} shows the effective spherical mass and concentration $m_{200} - C_{sph}$ contours, and their 1d marginalized parameter distributions, obtained fitting a triaxial NFW under all priors to the prolate and oblate  Line of Sight cases.  The contours and parameter distributions appear very similar to their fully triaxial counterparts, and exhibit the same trends of size and accuracy.


None of the contours or 1d parameter distributions for the Prolate LoS case include the true value of the underlying halo in either the triaxial or effective spherical parameterisation.  To better understand this, Figure \ref{fig:plot6} plots the full posterior probability distribution of the triaxial model under a Flat prior, and Figure \ref{fig:plot7} plots several 3D plots, colouring the $M_{200}-C$ contours and the $\theta-\phi$ contours by the axis ratios and, for the first pairing, orientation angles.  These allow better understanding of what degeneracies exist between the parameters, and thus what parts of parameter space are opened by allowing extreme axis ratios or particular halo orientations.  

First, we see that the failure of even the very general Flat prior contours to enclose the true halo value is due to the intrinsic priors on the orientation angles, which make the Line-of-Sight orientation of the lensing halo highly unlikely (for this lensing configuration $\theta=0$, which under random orientation has probability approaching zero).  This can be understood intuitively by imagining oneself at the centre of this ``rugby-ball" halo; there are only two places on the sky above where observers can look straight ``down the barrel'' of the halo -- those two special points situated directly at the pointed ends of the ball.  Conversely, there are many places on the sky where observers can take in identical side views of the halo -- all the many points radiating out from the circumference of the fattest part of the ``ball.''  

However, despite its statistical unlikelihood, this geometry is still important as a limiting case, as it is the strongest lensing configuration possible for a halo of a given mass and minor axis ratio.  Thus, though halos in such orientations are uncommon, they are strongly favoured in lensing-selected samples.  Further, they are particularly dangerous in that they show little (in the symmetric limit shown here -- no) ellipticity on the sky, and so are likely to be treated as spherical if triaxial modelling is used only selectively.  In some analyses of very powerful lenses, a more complex prior, taking into account lensing efficiency in the prior distribution of axis ratios and orientation angles, may be required to capture the true posterior distribution.  Looking at panels 1-3 in Figure \ref{fig:plot6} highlights this, as we see that it is low axis-ratio, low $\theta$ (close to Line of Sight), low mass solutions that give the models closest to the true lens: those that would be favoured by a strong lensing efficiency prior.  Figure \ref{fig:plot8} shows the resulting contours in the completely unrealistic scenario in which the true underlying axis ratios are known, here set to be $a=b=0.4$, while the mass, concentration, and orientation angles remain free.  The contours now contain the true model values for mass and concentration, confirming that the triaxial fitting routine behaves predictably and correctly in limits of both maximum and minimum knowledge about the underlying lens geometry.

\begin{figure*}
\epsfig{file=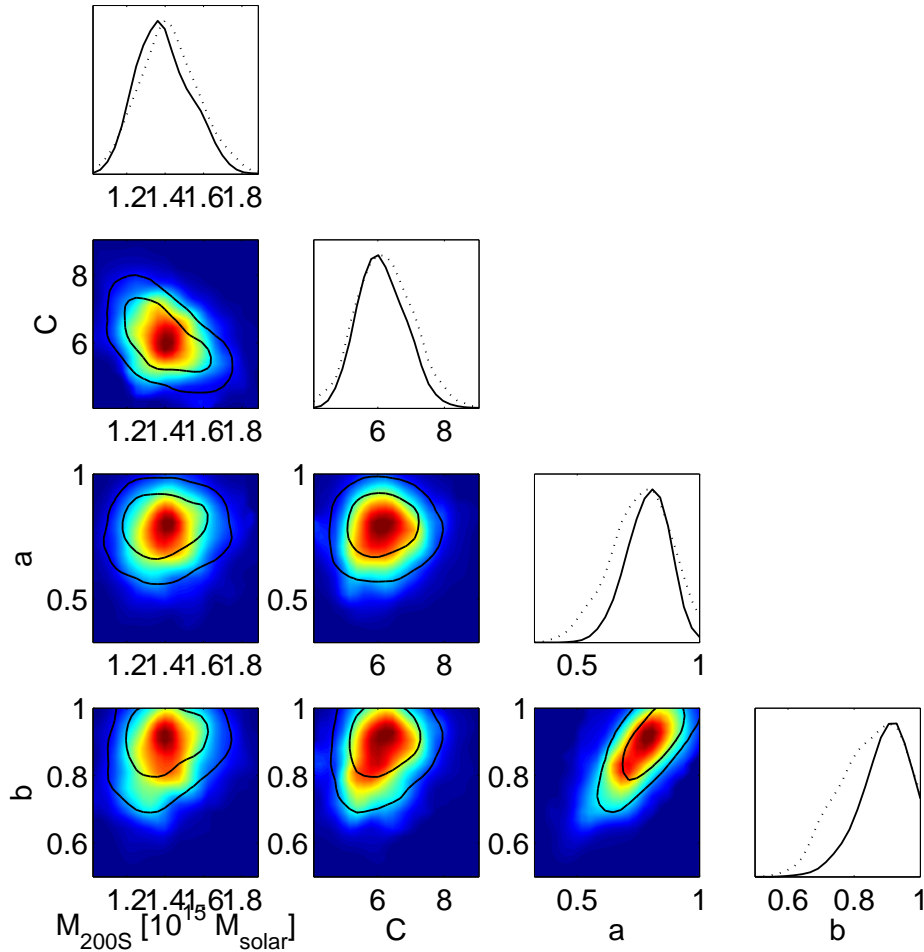,scale=0.7}
\caption{The full posterior probability distribution for a triaxial NFW fitted to weak lensing by an extreme prolate halo {a=b=0.4} in a Line of Sight orientation, under a Shaw prior.  The black contours give the 68$\%$ and 95$\%$ confidence contours of the marginalized posterior probability distribution, and the shading shows the marginalized likelihood values, smoothed over 20 bins (rather than 40 as in other figures) for increased clarity.  The solid (dotted) lines in the unshaded distributions at the end of each row give the 1-parameter marginalized probability (likelihood) distributions.}
\label{fig:plot9}
\end{figure*}

\begin{figure*}
\epsfig{file=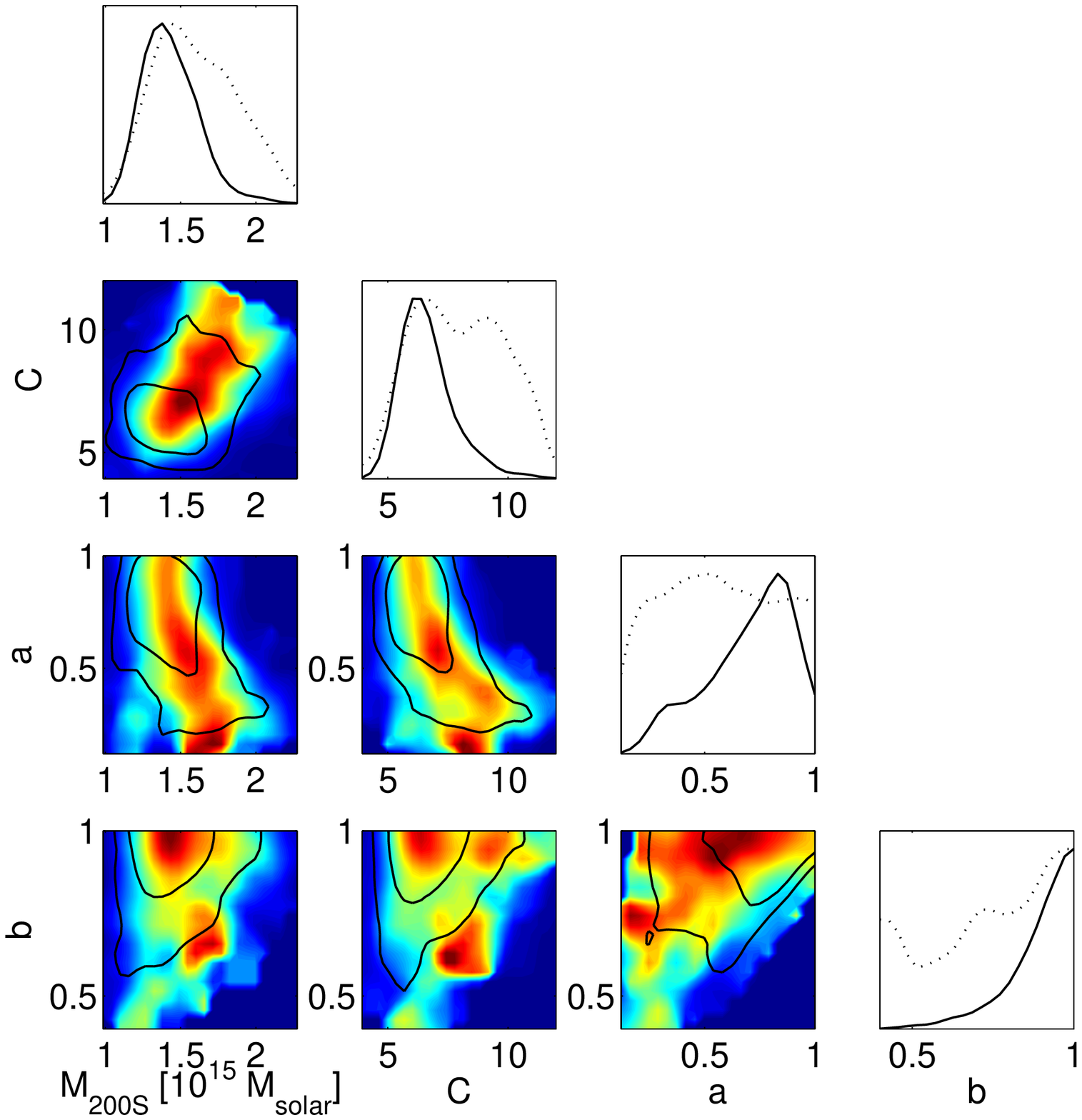,scale=0.7}
\caption{The full posterior probability distribution for a triaxial NFW fitted to weak lensing by an extreme prolate halo {a=b=0.4} in a Line of Sight orientation, under a Mass prior.  The black contours give the 68$\%$ and 95$\%$ confidence contours of the marginalized posterior probability distribution, and the shading shows the marginalized likelihood values, smoothed over 20 bins (rather than 40 as in other figures) for increased clarity.  The solid (dotted) lines in the unshaded distributions at the end of each row give the 1-parameter marginalized probability (likelihood) distributions.}
\label{fig:plot10}
\end{figure*}

Figures \ref{fig:plot9} and \ref{fig:plot10} show the posterior probability distribution for the triaxial model under the Shaw and Mass priors, omitting the orientation angles for brevity.  As expected, the axis ratio distribution is significantly constrained under the Shaw prior; this leads to tighter $M_{200}-C$ contours that overall favour lower masses and concentrations, but are completely contained within those obtained under the Flat prior. This indicates that some very low mass, low concentration models have also been lost.  Clearly, the Shaw prior, which has very little probability in the region where this halo actually sits in parameter space, is not a good prior to impose in this case.  The Mass prior, while weak, does move the contours in the direction of the true model, eliminating some higher mass cases while keeping the very low mass, low concentration cases.  This suggests that in cases in which there are reasons to suspect a particularly advantageous lensing geometry -- perhaps because a studied lens is one of the strongest in our observable universe, or dynamical studies suggest elongation along the line of sight -- a prior favouring lower masses may be an elegant and clear way to account for that prior belief, in place of a direct lensing efficiency prior.

When applying this MCMC triaxial NFW fitting method (or any other method) to real lensing data from galaxy clusters, there may be some uncertainty in the exact location of the cluster centre.  This may be accounted for if necessary by including the centre position of the halo model as an unknown in the fit, increasing the number of free parameters by two for all models.  However, we find that small errors in the fixed position of the cluster centre lead to only small changes in the most probable parameters recovered by our MCMC method, and that those changes are of the same scale under all priors, including at the extremes the highly triaxial Flat prior and the over-constrained Spherical prior.  The errors induced by fixing the cluster centre are therefore no more a problem in a triaxial analysis than in any other, and we neglect them herein in our assessment of the behaviour of our new triaxial NFW fitting method.

\begin{figure*}
\epsfig{file=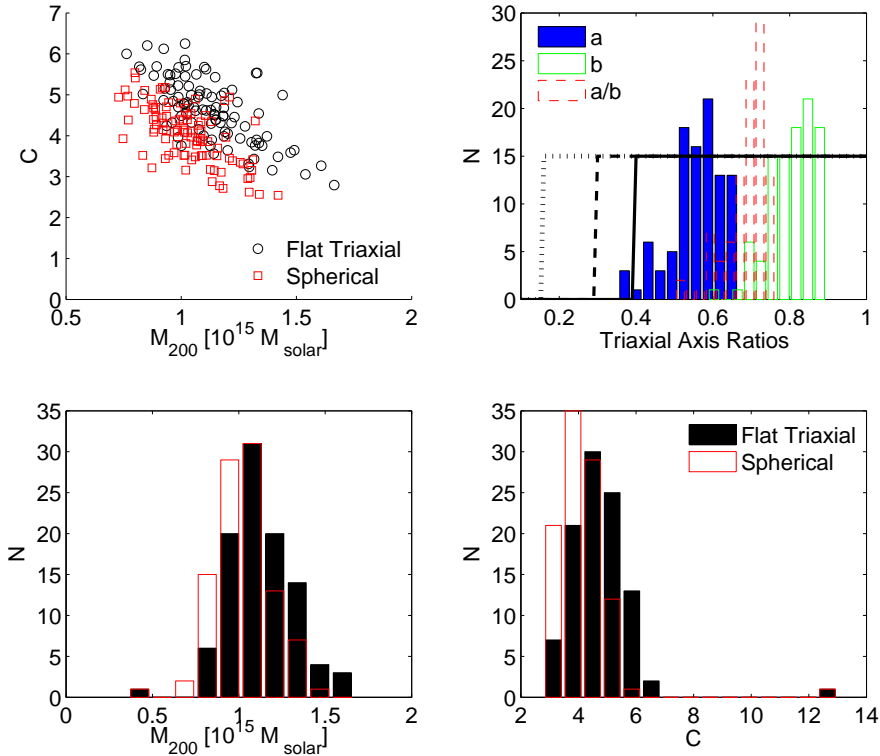,scale=0.6}
\caption{Parameter distributions resulting from fitting the triaxial NFW to a randomly oriented population of 100 lensing halos with $M_{200}=10^{15}$ M$_{\odot}$, $C=4$ and axis ratios drawn from the Shaw prior.  The results under the very general Flat prior are compared to those under a Spherical prior.  The top left panel plots the mean best-fitting concentration and mass values for each lens in the population under the two priors; the top right panel plots the returned axis ratio distributions (histograms) under the Flat prior compared to the prior distributions themselves (p(b) solid line, p(a/b) dashed line, p(a) dotted line); the bottom panels plot the distributions of mass and concentration obtained under each prior.}
\label{fig:plot11}
\end{figure*}

\begin{figure*}
\epsfig{file=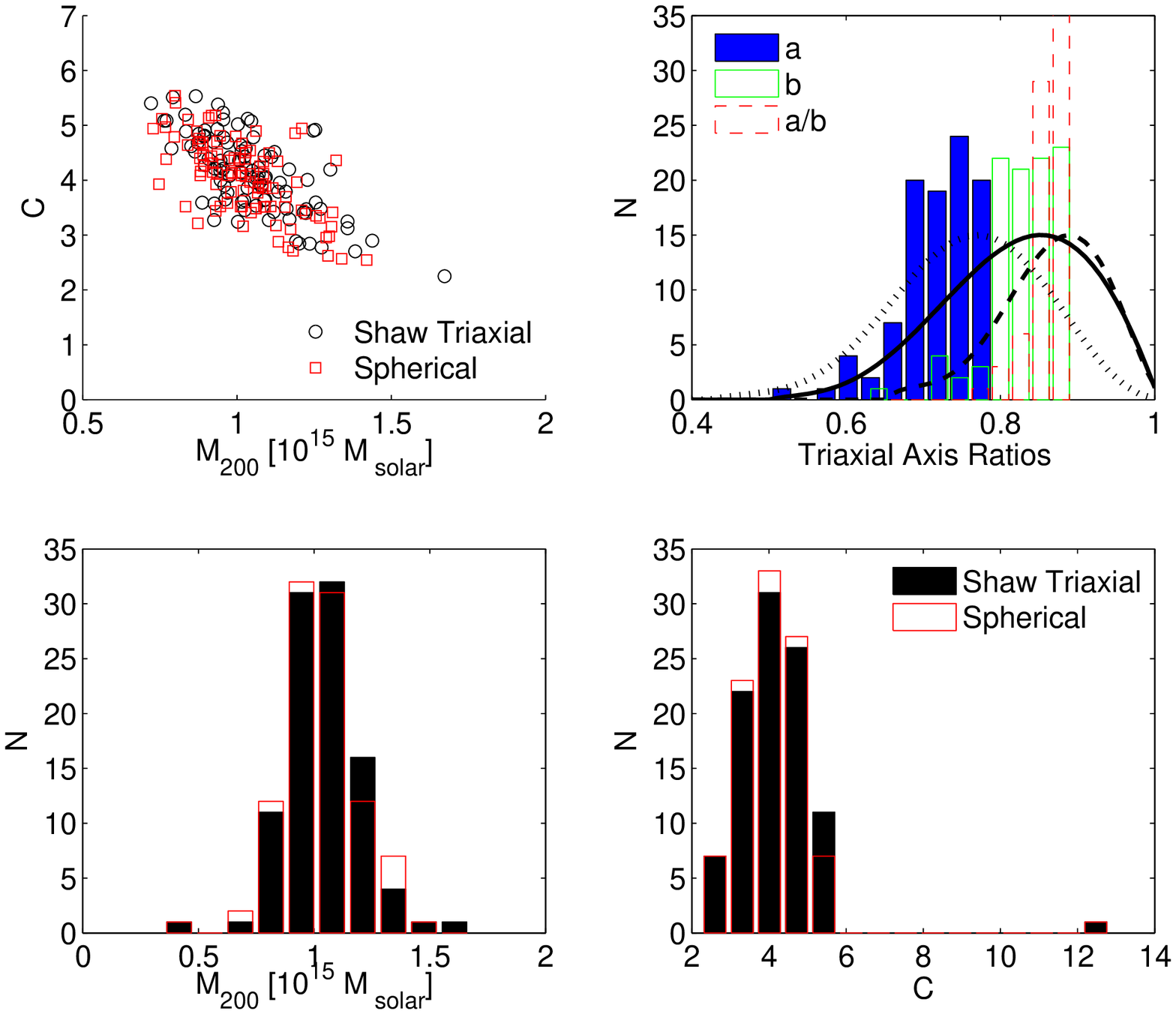,scale=0.6}
\caption{Parameter distributions resulting from fitting the triaxial NFW to a randomly oriented population of 100 lensing halos with $M_{200}=10^{15}$ M$_{\odot}$, $C=4$ and axis ratios drawn from the Shaw prior.  The results under the Shaw prior (which matches the distribution of the lens population) are compared to those under a Spherical prior.   The top left panel plots the mean best-fitting concentration and mass values for each lens in the population under the two priors; the top right panel plots the returned axis ratio distributions (histograms) under the Shaw prior compared to the prior distributions themselves (p(b) solid line, p(a/b) dashed line, p(a) dotted line); the bottom panels plot the distributions of mass and concentration obtained under each prior.}
\label{fig:plot12}
\end{figure*}

\subsection{Fitting to a population of halos}\label{ssec:pop}
We now move from looking at the behaviour of the triaxial NFW fitting technique on individual, highly triaxial halos, to its behaviour across a physically-motivated population of triaxial halos.  We use a standard Monte Carlo technique to choose 100 triaxial NFW halos with $M_{200}=10^{15}M_{\odot}$ and $C=4$ and axis ratios drawn from the distributions in \cite{shaw}.  The parameter distributions of the population are shown in Appendix \ref{sec:appc}; the mean effective spherical parameters for the population are similar to the constant triaxial values: $<m_{200}> = 0.99 \times 10^{15}M_{\odot}$ and $<C_{sph}> = 3.98$.  We randomly orient the hundred halos, lens through them as described in \S\ref{sec:simulations}, and carry out MCMC fits of the triaxial NFW model under the various priors previously described.  Figure \ref{fig:plot11} shows the resulting most-probable triaxial parameter distributions for the population under a Flat prior and compares them with a Spherical prior typically employed in lensing analyses. Figure \ref{fig:plot12} shows the same results under a Shaw prior, which exactly matches the true shape distribution of the population.

Under the Flat prior, the distribution of mean best-fitting parameters is skewed generally towards higher values of concentration and mass compared to that under the Spherical prior. Despite the underconstrained nature of the problem, the recovered axis ratios show a significant dependence on the  lensing data, as uniform priors for the axis ratio distributions 
become distributions peaked towards higher values of $a$ and $b$. 
Under the Shaw prior, the triaxial models give mean best parameter values very similar to those under the Spherical prior.  This reflects the similar shapes of the posterior probability confidence contours (e.g. the similar 'banana'-shaped $M-C$ contours) under these two priors.

The upper limits of the intermediate and minor:intermediate axis ratios are strongly constrained by the visible projected ellipticity on the sky ($b$ or $a/b$ must be less than or equal to $q$), while a soft lower limit is imposed by the fact that there are only a few geometries that allow for axis ratios significantly less than the observed ellipticity to be hidden in projection. 

Note that throughout we quote mean best-fitting parameter values as opposed to peak probability values; this gives parameter values further away from the extremes of the distribution. This is seen, for example, looking back at the 1D parameter distributions for the extreme halos in Figure \ref{fig:plot4}; although the probability distribution for $b$ peaks at $b=1$ for both the Prolate and Oblate LoS cases, the mean best-fitting values are $b=0.89$ and $b=0.88$ respectively.

\begin{figure}
\epsfig{file=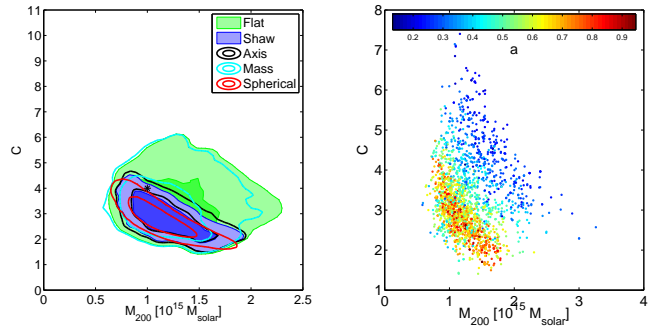,scale=0.3}
\caption{The left panel shows the $M_{200}-C$ 68$\%$ and 95$\%$ confidence contours, under various priors, for a triaxial NFW fit to a triaxial halo typical of those found in simulations, with $M_{200}=10^{15}M_{\odot}$ and $C=4$, $\{a=0.76, b=0.85\}$.  The right panel plots the $M_{200}-C$ distribution shaded by the minor axis ratio $a$ for the same halo under a Flat prior.}
\label{fig:plot13}
\end{figure}

Figure \ref{fig:plot13} shows $M_{200}-C$ confidence contours for a typical triaxial halo from the population, with axis ratios $\{a=0.76, b=0.85\}$.  Note that opening up the axis parameter space generally leads to many models with high mass and high concentration compared to the over-constrained spherical model, and only a few lower mass, lower concentration models.  This is the general behaviour of the posterior probability distribution for all halos, and indicates that neglected triaxiality in lens models cannot explain a whole population of massive halos with high concentrations within the $\Lambda$CDM paradigm.  It is only in a small fraction of orientations that highly triaxial halos lead to highly efficient lensing and thus to overestimates of mass and concentration under a spherical prior.  This is seen in the second panel of Figure \ref{fig:plot13}, plotting the $M_{200}-C$ contours under the Flat prior coloured by minor axis ratio $a$.  Most of the small-$a$ models are in the upper right corner of the distribution, but there is also a narrow band of them forming the lower left boundary of the posterior distribution, representing the highly efficient triaxial lens orientations that lead to overestimates of mass and concentration under standard spherical analysis methods.  Further note that the familiar degeneracy between mass and concentration is still evident for each range of minor axis ratio, with the allowance of more extreme axis ratios adding additional parallel lines of likely models, as indicated by the colour banding.

\begin{table}
\centering
\caption{The percentage of triaxial NFW fits to a population of 100 lensing halos (with $M_{200}=10^{15}$ M$_{\odot}$, $C=4$ and axis ratios drawn from the $\Lambda$CDM-motivated Shaw axis ratio distribution) for which the true mass and concentration values fall within the 68$\%$ and 95$\%$ confidence contours, under various priors on the halo parameters.}
\label{table1}
\begin{tabular}[t!]{ccccccccccccc}
\hline
Prior&68$\%$&95$\%$\\
\hline
Flat&86&99\\
{\bf Shaw}&{\bf 66}&{\bf 94}\\
Axis&70&96\\
Mass&86&99\\
Spherical&53&81\\
\hline
&Effective Spherical Parameterisation&\\
Flat&84&99\\
Shaw&61&89\\
Spherical&56&82\\
\hline
\end{tabular}
\end{table} 

The top half of table \ref{table1} gives the number of cases in which the true triaxial model value falls within the $68\%$ and $95\%$ confidence contours under each prior.  The bottom half of the table gives the equivalent results for the Flat, Shaw, and Spherical prior using the effective spherical parameterisation for both the fitted triaxial models and the triaxial lenses they are fit to.  Most strikingly, under the triaxial parameterisation the numbers correspond very well to the predicted statistics when the correct prior is employed; under the Shaw prior indeed 66 of the halos include the true value in their $68\%$ contour, and 94 include the true value in their $95\%$ contour!  This is very good evidence that the MCMC fitting technique behaves statistically as expected, in that the errors under a correct prior are the true errors due the full uncertainties in the problem: the statistical uncertainty in the shapes of the background galaxies, and the geometric uncertainty regarding the line-of-sight structure of the lensing halo.  As expected, the very general Flat prior gives larger contours that in a larger number of cases contain the true values. The only slightly stronger Mass prior behaves similarly, and the heavily over-constrained Spherical prior gives small contours that far too often exclude the true parameter values.

Since we will never know the exact axis ratio distribution of the galaxy cluster population, we test the response of the method applied to the same Shaw population of halos using a slightly different physically motivated prior taken loosely from \cite{bett}, plotted in Figure \ref{fig:plot1} and called herein the Axis prior. We define it simply by two normalised Gaussians, with $p(b)$ having mean $\mu = 0.8$ and standard deviation $\sigma = 0.125$ and $p(a/b)$ slightly narrower with $\mu = 0.85$ and $\sigma = 0.1$.  The result is encouraging; it is almost as good an estimator of both the mean parameter values across the population and the error contours on those parameters.  Thus, a slight mismatch between physically motivated priors and the true axis ratio distribution will not significantly decrease the accuracy and usefulness of this method.

Using the effective spherical parameterisation, fits employing the triaxial Shaw prior still statistically outperform the over-constrained Spherical prior, demonstrating that the triaxial model with a well-chosen prior better recovers even effective spherical parameters than does an over-simple spherical model.  However, fits under the Shaw and Axis priors using the full triaxial parameterisation are the overall best statistical performers, supporting our choice to use the fully triaxial parameterisation whenever possible.

The top half of Table \ref{table2} gives the mean triaxial mass and concentration values recovered across the population (fiducial values $M_{200} = 1.0\times10^{15} M_{\odot}$ and $C=4$); the bottom half gives the means of their effective spherical counterparts under the Flat, Shaw, and Spherical priors (fiducial mean values $m_{200} = 0.986\times10^{15} M_{\odot}$ and $C_{sph} = 3.98$).  Interestingly, and in line with earlier findings in CK07, the spherical model serves, across this realistic triaxial halo population, as a very good estimator of the mean triaxial and effective spherical mass and concentration.  While the fact that so often the contours under the Spherical prior do not contain the true parameter values makes it clear that the model is a bad choice for individual halos (it does not accurately constrain mass and concentration simultaneously, and the error estimates it provides are signficantly too small), this result suggests that if only mean values are needed, the spherical model may be adequate.
 
However, in the context of measuring a mass function or under poorer observing conditions where the error contours are more asymmetric, its inability to constrain individual halo models or their errors accurately make the spherical model most likely inadequate. This question will be further investigated in an upcoming paper by the authors.  As expected, the triaxial fitting method under the Shaw prior is also a good estimator of the mean population values in both the triaxial and effective spherical parameterisations, and crucially also estimates the errors accurately. The Flat prior and the closely related Mass prior result in overestimates of mass and concentration, as expected given the shape of the posterior probability distribution (many high-$M$, high-$C$ models).

\begin{table}
\centering
\caption{The mean most-probable parameter values resulting from fitting a triaxial NFW to a population of 100 lensing halos (with $M_{200}=10^{15}$ M$_{\odot}$, $C=4$ and axis ratios drawn from the $\Lambda$CDM-motivated Shaw axis ratio distribution), under various priors on the halo axis ratios and mass.}
\label{table2}
\begin{tabular}[t!]{ccccccccccccc}
\hline
Prior&$M_{200} [10^{15}M_{\odot}]$&$C$\\
\hline
Original Population&1.00&4.0\\
Flat&1.12&4.7\\
Shaw&1.04&4.2\\
Axis&1.04&4.3\\
Mass&1.07&4.7\\
Spherical&1.02&4.1\\
\hline
Effective Spherical Parameterisation&$m_{200}$&$C_{sph}$\\
\hline
Original Population&0.986&3.98\\
Flat&1.06&4.6\\
Shaw&1.03&4.2\\
Spherical&1.02&4.1\\
\hline
\end{tabular}
\end{table} 
\section{Discussion \& Conclusions}

Some may argue that it is foolhardy to fit a density profile model that fundamentally cannot be fully constrained with available data. 
Indeed, in many cases calculating the Bayesian evidence would likely 
favour a Spherical over a Flat prior, due to a significant decrease in available parameter space coupled with a relatively small decrease in likelihood values under the Spherical prior. 
However, unlike in cases of fitting multiple halos to account for substructure, or in another context, adding additional parameters to cosmological models, we know {\it a priori} from physical observations that a triaxial model is a better model than a spherical model -- we see clearly in non-parametric lensing mass maps that galaxy clusters have significant levels of ellipticity and are not spherical.  The parameter estimates and uncertainties resulting from spherical models, or 2D elliptical models, that assume a $\delta$-function prior on at least one axis ratio, therefore intrinsically claim a level of certainty we know we do not possess.  Making our best choice of prior on the axis ratios and fitting triaxial models is the more physically true approach, and gives the best estimate of parameter values and errors that can be obtained from lensing data alone.  The large size of these contours emphasises the need to combine lensing data with other data sources, be they SZ, dynamical, or X-ray, because the  dependence of all lensing effects -- strong, weak, intermediate -- on projected mass fundamentally constrains the precision and accuracy with which galaxy cluster masses and concentrations can be obtained.

\begin{figure}
\epsfig{file=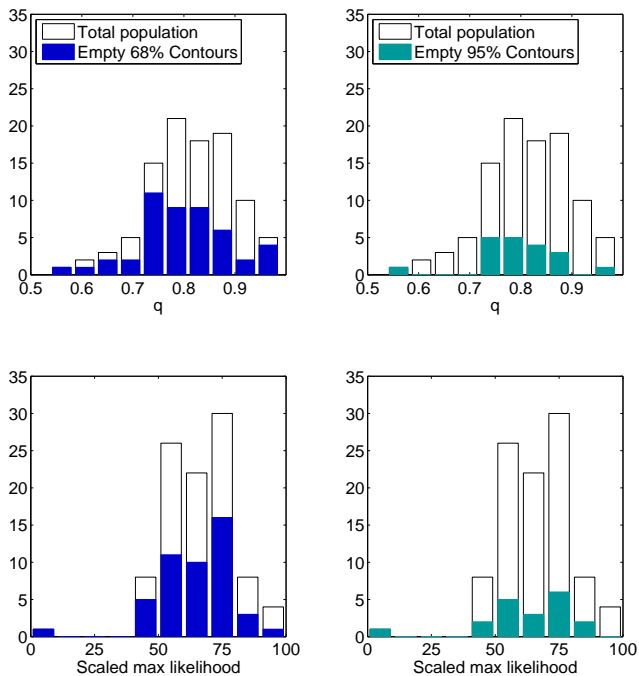,scale=0.4}
\caption{The right (left) top and bottom panels show, respectively, the distributions of the projected axis ratios $q$ and the maximum likelihood values obtained from fitting a spherical NFW, for the halos of the triaxial lensing population of Section \ref{ssec:pop} for which the 68\% (95\%) confidence contours recovered fitting an NFW under a Spherical prior {\it do not} enclose the true triaxial halo values.}
\label{fig:plot14}
\end{figure}

Further, there is no way to know for a single lensing cluster how important triaxiality is without first carrying out a full triaxial analysis and comparing its results to those from a spherical fit; Figure~\ref{fig:plot14} helps illustrate why this is the case.  The four panels plot the distributions of two observable properties of the triaxial halos of the lensing population of Section \ref{ssec:pop} for which the 68\% and 95\% confidence contours obtained fitting an NFW under a Spherical prior {\it do not} contain the true halo values.  These observables are the axis ratio of the projected isodensity contours $q$ of the lens and the maximum likelihood value of the NFW fit under a Spherical prior.  It might be hoped that either a high visible ellipticity (low $q$) or a low likelihood value would indicate that the spherical model is a bad fit; however, as seen in the figure, the distributions show no bias compared to the distribution of the entire lensing population, and thus can provide no indication of the adequacy of the spherical fit.  This can be understood by example by referring back to the extreme halo shapes and orientations of Section \ref{ssec:ex}: the highly triaxial prolate halo in a Line of Sight orientation will have both a low ellipticity ($q\sim 1$) and a very high maximum likelihood value (it is an efficient lens with circular projected isodensity contours and so will be very well fit by a spherical model), but the most-probable parameter values and their errors obtained from fitting a spherical NFW are very inaccurate.  Thus, it is necessary to fit a triaxial NFW using the best prior available to every lens, as there is no way to determine beforehand how triaxial any given halo might be.

A great advantage of this MCMC method is the ease with which other data sets and constraints may be incorporated.  Whether through extra terms in the likelihood function (though such inclusions require a careful analysis, not seen yet in the literature for the combination of many data types, of the proper weights to give data types with very different scales and sources of errors), or through a prior (e.g. a prior on the cluster mass taken from the best estimate and errors of a dynamical study). 

The future of understanding cluster structure lies in the combination of various data types, as all available methods are limited either by fundamental constraints, such as lensing, or by the availability of data, as in dynamics, in which highly simplified dynamical models are required by the relatively small numbers of position-velocity pairs.  Methods such as this, and those currently being developed by e.g., \cite{jullo} and \cite{feroz}, are a crucial element of the next generation analysis toolbox for galaxy cluster studies.

\subsection{Selecting better priors}
Placing well-motivated priors on the shapes of galaxy clusters is crucial to future work modelling the most massive structures in the universe, and understanding both their characteristics as individuals and a population.  While structure formation simulations provide a good starting point for such priors, using their predictions does of course bias all results towards agreement with those simulations.  One good alternative is to use a distribution of 2d axis ratios observed in a sample of galaxy clusters, preferably from mass-sensitive methods (e.g. lensing mass reconstructions) and from this construct a 3d shape prior.  For this, even the simplest elliptical lensing models would be adequate, as CK07 showed that even a Singular Isothermal Ellipsoid model consistently recovered projected axis ratios accurately, even if the true lensing profile was NFW (this is not true for other model parameters).

While the necessity and importance of these priors may be discomforting, using standard techniques and simpler models is simply disguising the problem, as such models contain highly-restrictive hidden priors.  Bayesian techniques such as the one presented in this paper are simply tools that allow us to better understand the true constraints we can place on physical models, not solutions in themselves to physical problems.  The broad posterior probability distributions for the triaxial NFW profiles fit to simulated lensing data in this paper indicate the weakness of our current constraints, and emphasise the need for focus on the careful combination of complementary data types to further constrain galaxy cluster structure models. To this end, this Bayesian MCMC triaxial NFW fitting  method provides, through the prior probability functions, a statistically robust and straightforward way to combine constraints from data types with very different error properties.

\subsection{Application}
This MCMC triaxial NFW fitting method, using fully tested look-up tables to significantly speed up the calculation of the triaxial NFW lensing quantities, can be implemented for a standard weak lensing data set on a single fast processor in about 1 day, making it fully feasible for use across even the largest existing lens surveys.  To begin this process, in a companion paper Corless, King, \& Clowe (in preparation), we apply this method employing a range of priors to galaxy clusters Abell 1689, Abell 1835 and Abell 2204.  

In applying this method to observational data it will be important to account for associated and line-of-sight structures in the lensing field.  Galaxy clusters form at the intersections of filaments, so it is not surprising that spectroscopic studies often reveal correlated structures in their fields(e.g. \cite{lokas}) that can bias our estimation of cluster parameters if neglected (\cite{kingb}).  Unaccounted for large scale structure along the line of sight can also have a severe impact on lensing analyses (\cite{hoekstra}),leading to a factor of $\sim 2$ increase in errors on cluster mass. As demonstrated by \cite{dodel}, this can be somewhat mitigated by including a noise term for large scale structure in the lensing analysis.  Building on the greater coherence scale of large scale structure noise compared with the noise associated with the intrinsic ellipticity dispersion of galaxies, Dodelson notes that the errors on cluster mass can be reduced by $\sim 50\%$ for wide-field data.

This MCMC triaxial fitting method will be very useful in the analysis of current and future surveys such as LoCuSS, PAN-STARRS and DES.

\subsection{Summary}
We here present an MCMC method for fitting triaxial NFW models to weak lensing data that
\begin{itemize}
\item consistently and accurately returns parameter and error estimates representing the true uncertainties of the problem;
\item includes Bayesian priors on the halo geometry, chosen from simulations or observations;
\item allows great flexibility in the inclusion of constraints from different observations and analyses, including strong lensing, dynamical studies, and S-Z and X-ray derived mass models;
\item demonstrates that while triaxiality can explain rare cases of overconcentrated galaxy cluster halos within the $\Lambda$CDM paradigm, it cannot explain a population-wide trend of over-concentration;
\item should be widely applied in weak lensing analyses to better determine cluster parameters and errors that represent the true (and limited) extent of our knowledge of the distribution of matter in galaxy cluster lenses.
\end{itemize}

\section*{Acknowledgments}
This work was supported by the National Science Foundation, the Marshall Foundation, and the Cambridge Overseas Trust (VLC) and the Royal Society (LJK).  We thank Antony Lewis, Matthias Bartelmann, and Hiranya Peiris for very useful discussions, and our anonymous referee for a thorough and highly constructive review of this work.  VLC also thanks ESO Chile and Chris Lidman for hospitality and support during the course of this work.

\appendix

\section{Lensing through triaxial halos}\label{sec:appa}

Following OLS, the triaxial halo is projected onto the plane of the sky to find its projected elliptical isodensity contours as a function of the halo's axis ratios and orientation angles ($\theta$, $\phi$) with respect to the the observer's line-of-sight.\footnote{Although we set $c=1$, we keep $c$ as a variable in our notation for consistency with OLS.}    The elliptical radius is given by
\begin{equation}\zeta^2 = \frac{X^2}{q_X^2} + \frac{Y^2}{q_Y^2}\label{eq:xi}\end{equation}
where $(X,Y)$ are physical coordinates on the sky with respect to the centre of the halo,
\begin{eqnarray}
q_X^2&=&\frac{2f}{\mathcal{A}+\mathcal{C} - \sqrt{(\mathcal{A}-\mathcal{C})^2 + \mathcal{B}^2}}\\
q_Y^2&=&\frac{2f}{\mathcal{A}+\mathcal{C} + \sqrt{(\mathcal{A}-\mathcal{C})^2 + \mathcal{B}^2}} \end{eqnarray}
where
\begin{equation}f = \sin^2\theta\left(\frac{c^2}{a^2}\cos^2\phi + \frac{c^2}{b^2}\sin^2\phi\right) + \cos^2\theta,\label{eq:f}\end{equation}
and
\begin{eqnarray}
\mathcal{A}&=&\cos^2\theta\left(\frac{c^2}{a^2}\sin^2\phi + \frac{c^2}{b^2}\cos^2\phi\right) + \frac{c^2}{a^2}\frac{c^2}{b^2}\sin^2\theta,\\
\mathcal{B}&=&\cos\theta\sin 2\phi\left(\frac{c^2}{a^2} - \frac{c^2}{b^2}\right),\\
\mathcal{C}&=&\frac{c^2}{b^2}\sin^2\phi + \frac{c^2}{a^2}\cos^2\phi.
\end{eqnarray}
The axis ratio $q$ of the elliptical contours is then given by
\begin{equation}q = \frac{q_Y}{q_X}\label{eq:q}\end{equation}
and their orientation angle $\Psi$ on the sky by
\begin{equation}\Psi = \frac{1}{2}\tan ^{-1}\frac{\mathcal{B}}{\mathcal{A}-\mathcal{C}}~~~(q_X \ge q_Y).\label{eq:psi}\end{equation}

Here we diverge slightly from OLS's treatment as we are interested not in deflection angles but in the lensing shear and convergence, both combinations of second derivatives of the lensing potential $\Phi$ (commas indicate differentiation):
\begin{eqnarray}
\gamma_1&=& \frac{1}{2}\left(\Phi_{,XX} - \Phi_{,YY}\right),\\
\gamma_2 &=&\Phi_{,XY},\\
\kappa &= &\frac{1}{2}\left(\Phi_{,XX} + \Phi_{,YY}\right).
\label{eq:gammakappa}\end{eqnarray}
These derivatives are calculated as functions of integrals of the spherical convergence $\kappa(\zeta)$ (see e.g. \cite{bartel} for a full treatment of weak lensing by a spherical NFW profile) following the method of \cite{schramm} and \cite{keeton}, normalised by a factor of $1/\sqrt{f}$ from Equation~\ref{eq:f} (see OLS for the derivation of this normalisation)
\begin{eqnarray}
\Phi_{,XX} &= &2qX^2K_0 + qJ_0,\\
\Phi_{,YY} &= &2qY^2K_2 + qJ_1,\\
\Phi_{,XY} &= &2qXYK_1,\end{eqnarray}
where
\begin{eqnarray}K_n(X,Y)& =&\frac{1}{\sqrt{f}}\int_0^1 \frac{u\kappa'(\zeta(u)^2)}{[1 - (1-q^2)u]^{n+1/2}}du,\label{eq:K}\\
J_n(X,Y)& =&\frac{1}{\sqrt{f}}\int_0^1 \frac{\kappa(\zeta(u)^2)}{[1 - (1-q^2)u]^{n+1/2}}du,
\end{eqnarray}
and
\begin{equation}\zeta(u)^2 = \frac{u}{q_X}\left(X^2 + \frac{Y^2}{1 - (1-q^2)u}\right).\end{equation}

Note that our radial variable $\zeta$ appears different from Keeton's $\xi$ because it is defined in terms of two axis ratios $q_X$ and $q_Y$ rather than one $q$: $\zeta = \xi/q_X$.  This reflects a dependence on the 3D structure of the cluster; for example, extended structure along the line of sight decreases $q_X$ and thus increases the convergence and shear at a given $(X,Y)$.
\begin{figure}
\epsfig{file=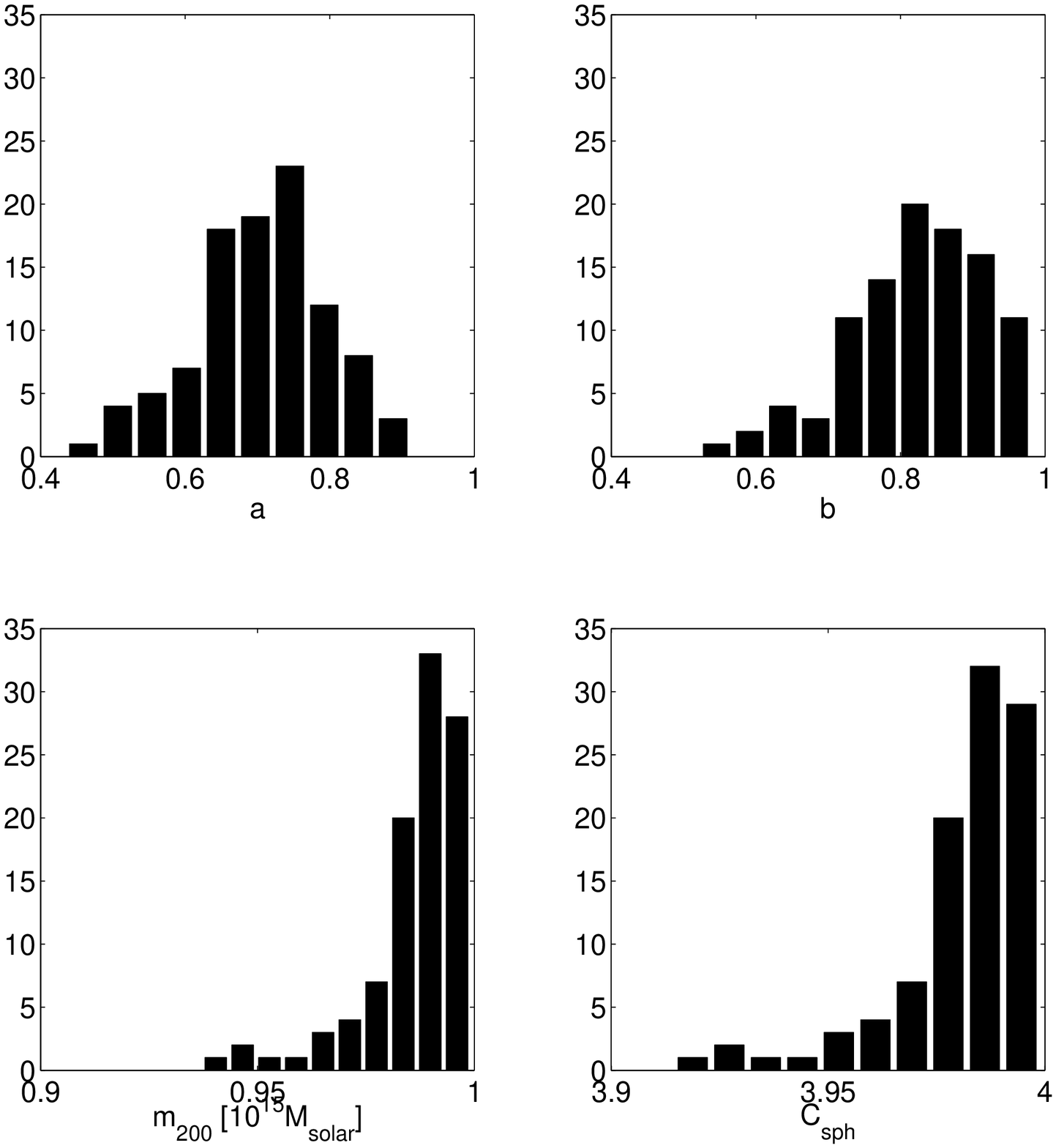,scale=0.4}
\caption{The distribution of axis ratios $a$ and $b$ and effective spherical mass $m_{200}$ and concentration $C_{sph}$ for the lensing population of Section \ref{ssec:pop}.}
\label{fig:plotA1}
\end{figure}

\section{Parameters of the Lensing Population} \label{sec:appc}
The 100 halos that make up the lensing population of Section \ref{ssec:pop} each have triaxial virial mass $M_{200} = 1.0\times 10^{15}$ M$_{\odot}$ and concentration $C=4.0$.  Their axis ratios are drawn from the distributions found in the structure formation simulations of \cite{shaw}, and their effective spherical virial masses $m_{200}$ and concentrations $C_{sph}$ are calculated numerically as a function of their triaxial shape.  The distributions of axis ratios $a$ and $b$ and the effective spherical parameters $m_{200}$ and $C_{sph}$ are plotted in Figure \ref{fig:plotA1}.

\onecolumn
\section{Priors from Simulations: Shaw Prior} \label{sec:appb}

We define the Shaw prior, plotted in Figure \ref{fig:plot1}, by fitting polynomials to the data points of Figure 14 in \cite{shaw}:

$p(b) = \left\{\begin{array}{ll} 0 & \textrm{if }b < 0.5\\\left[1.6329b^5 - 7.9775b^4+9.3414b^3 - 6.6558b^2+2.2964b - .3088\right]\times 10^3 & \textrm{if }0.5 \leq b \leq 1.0\end{array}\right.$

and

$p\left(\frac{a}{b}\right) = \left\{\begin{array}{ll}0& \textrm{if }\frac{a}{b} < 0.65\\
\left[5.76647\left(\frac{a}{b}\right)^6 - 2.459265\left(\frac{a}{b}\right)^5 + 42.3154\left(\frac{a}{b}\right)^4 - 37.2765\left(\frac{a}{b}\right)^3\right.&\textrm{if }0.65 \leq \frac{a}{b} \leq 1.0.\\
\left.~~~~~+ 17.4650\left(\frac{a}{b}\right)^2 - 4.00238\left(\frac{a}{b}\right)+ .32462\right]\times 10^4&\end{array}\right.$

\bsp

\label{lastpage}

\end{document}